# A two-step approach to account for unobserved spatial heterogeneity[1]

Anna Gloria Billé [a]*, Roberto Benedetti [b], Paolo Postiglione [b]

[a] Department of Economics and Finance, University of Rome Tor Vergata

[b] Department of Economic Studies, University G. d'Annunzio of Chieti-Pescara.

*corresponding author. Email: anna.gloria.bille@uniroma2.it

**Abstract.** Empirical analysis in economics often faces the difficulty that the data are correlated and heterogeneous in some unknown form. Spatial econometric models have been widely used to account for dependence structures, but the problem of directly dealing with unobserved spatial heterogeneity has been largely unexplored. The problem can be serious especially if we have no prior information justified by the economic theory. In this paper we propose a two-step procedure to endogenously identify spatial regimes in the first step and to account for spatial dependence in the second step, with an application to hedonic house price analysis.

## 1. Introduction and literature review

It is a well-established fact that if spatial models are correctly specified then they can also be consistently and efficiently estimated by the commonly used estimators. However, incorrect functional forms, correlated omitted variables, models with near unit roots and row-normalized weighting matrices, and so on, typically produce spurious spatial autocorrelations (Fingleton, 1999; McMillen, 2003, Lauridsen and Kosfeld, 2006; Lee and Yu, 2009; Lauridsen and Kosfeld, 2011), which can lead to inconsistency of the usual estimators. Spatial heterogeneity is a particular form of heterogeneity, usually unobserved, that is related to geo-referred data sets and would lead to misspecification of the model if not account for.

---

[1] The title of the first version of the paper was: "Spatial Heterogeneity in House Price Models: An Iterative Locally Weighted Regression Approach".

Empirical analysis in economics often faces the difficulty that the data are correlated and heterogeneous in some unknown form. A first attempt to explicitly model discontinuities in space is for example the work by McDonald and Owen (1986), which procedure has been then used by McMillen (1994) to study potential discontinuities in the population density of Chicago in 1980. As Anselin (1988a, p. 119) stressed, there are two distinct aspects that pertain to spatial heterogeneity: the former is the structural instability as expressed by changing functional forms or varying parameters, the latter is the heteroscedasticity which follows from missing variables or other forms of misspecification that lead to error terms with non-constant variance. In this paper, we are going to deal with the idea that coefficient estimates can vary over space leading to a spatial structural instability, i.e. when its parameters take on distinct values in subsets of the spatial sample. Moreover, if spatial heterogeneity can be categorized into a small number of regimes, each represented by different values for the regression coefficients, the phenomenon is also known as spatial regimes. In this case, if spatial heterogeneity is present the functional form of the model will be misspecified because of the wrongly assumed constant relationships between dependent variables and regressors. The following Anselin's (2010, p. 5) statement is useful to understand

> *"Spatial heterogeneity becomes particularly challenging since it is often difficult to separate from spatial dependence. This is known in the literature as the inverse problem. It is also related to the impossible distinction between true and apparent contagion. The essence of the problem is that cross-sectional data, while allowing the identification of clusters and patterns, do not provide sufficient information to identify the processes that led to the patterns."*

The problem of spatial heterogeneity in terms of spatially varying parameters has been largely unexplored by spatial econometricians, typically because of the main purpose of controlling only for spatial spillover effects. As Postiglione et al. (2013, p. 171) stressed "*the problem of spatial* *heterogeneity is often neglected in empirical analysis of geographic data and this negligence can affect sensibly model estimates*". Some authors have attempted to detect the presence of spatial heterogeneity by constructing statistical tests that are typically based on the LM statistic (Anselin, 1988b; de Graaff *et al.*, 2001; Lauridsen and Kosfeld, 2006; Lauridsen and Kosfeld, 2011; Pede *et al.*, 2014; among others). Unfortunately, once detected the presence of spatial heterogeneity no test is able to suggest how to correctly model our spatial data set and in which direction we have to proceed for further analyses. Recently, Ibragimov and Müller (2010) have derived the small and large sample properties of the *t* statistic, also in the context of spatially correlated data, by assuming a reasonable partition in *q* groups of the data. As Ibragimov and Müller (2010, p. 454) emphasized "*some a priori knowledge about the correlation structure is required ...*". However, in practical cases, there is usually no reason to accept one partition instead of another, which is in some way justified by the economic theory.

Following a parametric approach, the typical starting point to estimate a spatial econometric model is usually based on the choice of a row-standardized spatial weighting matrix, $W$, which is able to specify the relationship between neighboring observations. In some cases the significance of the spatial spillover effects through the autoregressive coefficient might be simply due to an *omitted spatially-correlated regressors problem*, which can easily justifies the use of the well-known more flexible spatial Durbin models (see e.g. Corrado and Fingleton, 2011, LeSage, 2014, for comprehensive discussions). However, neighborhood influence is not calibrated in terms of the data but is prescribed by the specification of $W$[2].

Imposing a predefined spatial structure of the data can be sometimes too restrictive in practical cases and it can bias results when inappropriate[3], so that McMillen (2012), among others, has criticized this approach. Although our purpose is not to criticize the parametric approach, it is reasonable to assume that for some economic phenomena there is no reason, justified by the economic theory, to choose a priori a particular spatial structure of the underlined spatial process. The main purpose of the present paper is then to propose a possible partition of the spatial data, i.e. a classification of the data due to unobserved heterogeneity, with no *a priori* information of the true dependence structure.

---

[2] See the papers of LeSage and Pace (2014), Getis (2009,2007) for considerations on the spatial weight matrix and the autocorrelation coefficient.

[3] Recently, both from a theoretical and a computational perspective, some excellent works on the definition of the W matrix has been proposed (see LeSage and Pace, 2007; Seya et al., 2013; Bhattacharjee and Jensen-Butler, 2013; Qu and Lee, 2015). In particular, Qu and Lee (2015) defined a particular endogenous W matrix (where the usual exogenous W matrix can be considered a particular case) and showed the consequences on the estimates by considering commonly used estimators in SAR cross-sectional models when the true W is endogenous.

For instance, the advanced recent literature in hedonic house price models accounts for spatial spillover effects but still ignore the possibility of a spatial heterogeneity effect (Holly et al., 2010; Holly et al., 2011). Some researchers are then recognizing that the spatial structures can be sufficiently different that the data should not be pooled and estimated together and global spatial regression models usually fail in taking into account any potential variations over space, with the consequences of biased resulting estimates. Along the same line, one of the purposes of the present paper is to show how the presence of spatial heterogeneity might modify (and generally reduces) the significance of the spatial effects in a global spatial autoregressive model by simply estimating spatial autoregressive models with spatial regimes (which introduce a sort of flexibility in spatial autoregressive models). This can be interpreted in the fact that the spatial heterogeneity might generates part of the spatial autocorrelation effect, or in other words that the autoregressive coefficient is sometimes overestimated. In line with this view is the work of Basile *et al.* (2014). In order to simultaneously account for spatial dependence, unknown functional form and unobserved heterogeneity, they proposed the use of the so-called Spatial Autoregressive Semiparametric Geoadditive Models, which are based on a combination between spatial parametric autoregressive models and unknown smooth functions.

Semiparametric and nonparametric estimation methods, such as geographically weighted regressions (GWRs) (Brunsdon *et al.*, 1996), are proving to be valid alternatives to parametric approaches and should be used as diagnostic tools to detect the presence of spatial heterogeneity[4]. The above methods allow coefficient estimates to vary over space by calibrating the global model separately for each spatial units in order to produce *n* sets of parameter estimates, with the behind basic idea that simple econometric models represent the data best in small geographic areas. They indeed have the attracting advantage to control for misspecified spatial effects while using highly flexible functional forms, with the only condition that nearby observations need more weight when constructing an estimate for a target point. McMillen and Redfearn (2010) showed that the GWR specification can be viewed as a special case of the already known locally weighted regression (LWR) method[5]. GWRs or Geoadditive Models usually interpret the spatially varying parameter problem as a smooth changing over space. As a matter of fact, pairs of beta coefficient GWR estimates that are proximal in space could not exhibit statistically significant differences. As Anselin (2010, p. 17) underlined *"...in models of spatial heterogeneity, the spatial regimes or spatially varying coefficients show evidence of the heterogeneity, but do not explain it. Ideally, one would want to make the structure of dependence and/or the structure of heterogeneity endogenous"*. This is the aim of the present paper. Our purpose is to iteratively find, in a way that spatial parameter variations can be described by breaks in continuity over space (or spatial regimes).

In the spatial statistics literature an adaptive weights smoothing (AWS) algorithm have been recently proposed by Polzehl and Spokoiny (2000, 2006) to describe, in a data-driven iterative way, a maximal possible local neighborhood of every point in space $i$ in which the local parametric assumption is justified by the data. The basic assumption of the proposed approach is that for every point $i$ there exists a vicinity of $i$ in which the underlying model can be well approximated by a parametric model with the constant set of parameters. Their method has been applied in the field of image analysis and it is based on both a successive increase of the local neighborhoods around every point $i$ and on a description of the local models by assigning weights to every spatial unit that depends on the result of the previous step of the procedure. The potential of this method in an econometric environment concerns the possibility of endogenously obtaining cluster of observations due to unobserved spatial heterogeneity (i.e. spatial regimes) exhibiting similar coefficient estimates.

---

[4] A valid alternative to the GWR approach is the NCSTAR model as pointed out by Lebreton (2005).
[5] The main difference pertains the way of thinking the distance measure. In GWRs, distance is thought as a mere geographic distance (Euclidean, great circle, etc.), whereas in LWRs an economic interpretation can be also assigned.

Andreano et al. (2016) defined a first algorithm based on the work by Polzehl and Spokoiny (2000, 2006) for the identification of economic convergence clubs. In this paper we substantially modify the previous contribution proposing a two-step procedure (see Section 3 for further details and comparisons), which is based on the conjunction between the LWR approach and the AWS procedure. A two-step procedure to deal with both spatial dependence and spatial heterogeneity within the estimation of hedonic house price functions has been proposed for instance by Beron et al. (2004). However, their method focused on the estimation of a first set of parameters related to environmental characteristics in the first step, whereas it proposed a spatial econometric model that accounts for both spatial dependence and spatial heterogeneity in the second step, assuming a spatial trend as a quadratic form of latitudes and longitudes for the heterogeneity effects. In our two step procedure, instead, we propose a first step that focuses on the estimation of unobserved discrete spatial heterogeneity (i.e. spatial regimes) and a second step that estimates the effects of both spatial dependence and the identified spatial regimes. Our purpose is based on the idea that we can combine the potential of local estimation with the usefulness of a modified AWS procedure, which is able to identify spatial regimes, i.e. subsamples over space with an estimated set of beta coefficients for each of them.

The paper is structured in the following way. In section 2 we introduce the LWR and the GWR as a special case. In section 3 we explain the first step of our procedure, *i.e.* the algorithm, whereas section 4 explain the second step to estimate both spatial dependence and spatial regimes. Section 5 illustrates the data set used and their main estimation results in terms of the marginal effects obtained by using different spatial econometric models. Finally, section 6 concludes.

## 2. Locally Weighted Regressions

Spatial econometric models may be not appropriate in the presence of (unobserved) spatial heterogeneity. Locally weighted regressions (LWRs) (Cleveland and Devlin, 1988) or Geographically weighted regressions (GWRs) (Brunsdon *et al.*, 1996; Fotheringham *et al.*, 1998, Fotheringham *et al.*, 2002), which are recognized to be natural evolutions of the expansion method (Casetti, 1972), allow us to estimate local rather than global parameters.

Residual terms usually exhibit a different from zero spatial autocorrelation parameter that, actually, might be not statistically different from zero if the true reason of error autocorrelation is different from a true contagion process (i.e. spurious autocorrelation). The first law of geography (Tobler, 1970, p. 236) states: "*everything is related to everything else, but near things are more related than distant things*". The goal should be to detect the reason why the closest things are related. The basic idea behind local geographical estimation is that simple linear functions may fit well for observations close to a site, say $i$, but probably they will be inappropriate when more distant observations are included. Limiting the estimation to a neighborhood of observation $i$ eliminates much of the heteroscedasticity and autocorrelation that is endemic to spatial data set. Therefore, simple linear functions can be written to account for local parameter estimates in the following way

$$y = (\beta \odot X)\mathbb{1} + \varepsilon, \quad \varepsilon \sim N(0, \sigma_\varepsilon^2 I) \qquad [2.1]$$

where $y$ is an $n$-dimensional column vector, $\beta$ is an $n$ by $p+1$ matrix with $i$-th row $\beta_i = (\beta_{i0}, \beta_{i1}, \beta_{i2}, \ldots, \beta_{ip})'$, $\odot$ is the Hadamard product operator in which each element of $\beta_i$ is multiplied by the corresponding element of the $i$-th row in $X$, $x_i = (x_{i0}, x_{i1}, x_{i2}, \ldots, x_{ip})$, $\mathbb{1}$ is a $(p+1)$ column vector of ones, and $\varepsilon$ is an $n$-dimensional column vector of innovations. The idea is that, in order to estimate a set of parameters for a given point $i$, one can approximate the model in [2.1] in the neighborhood of $i$ by using a simple linear model, and perform least squares with a subset of points that are closed to $i$. The objective function, for each local regression $i$, takes the following form

$$Q(\beta_i) = [y - X\beta_i]'W_i[y - X\beta_i] = \varepsilon'W_i\varepsilon \quad [2.2]$$

where $W_i$ is an *n*-dimensional diagonal square matrix whose diagonal elements, $(w_{i1}, w_{i2}, \ldots, w_{ij}, \ldots, w_{in})$, denote the weights (in our case geographic distances) of each of the *n* observed data for regression point *i*. We will therefore have *n* diagonal spatial weighting matrices and *n* sets of local parameter estimates that correspond to the local marginal effects. Then, a locally weighted least squares (LWLS) estimator is simply obtained by repeated weighted least squares

$$\hat{\beta}_i = (X'W_iX)^{-1}X'W_iy, \ i = 1, \ldots n. \quad [2.3]$$

Initial Kernel weighting functions have to be specified to define the $W_i$ matrices. Such functions are typically the exponential, the Gaussian, the bi-square, the tri-cube, and they seem to be similar in placing higher weight on nearby observations and lower weight on the distant ones, but they differ according to the type of bandwidth (*b*).

The choice of the bandwidth value is then crucial: the neighborhood of observation *i* is defined by the value of *b* which then determines the number of observations (i.e. a subsample) that receive a weight in constructing the estimate for *i*, and how rapidly these weights decline as distance increases. So, an important question could be how close to *i* should points be considered, or in other words in which way we can define a proper neighborhood for *i*. The *n* observations are used several times with a maximum equals to $n^2$. Every local $\hat{\beta}_i$ can be based on the same set of *n* observations, but in this case the local estimates would be the same. The minimum number of observations to be used for each local estimation is instead $p + 1$, but then the $\hat{\beta}_i$'s would be not significant. The point is then to choose an ideal subsample of observations for each local estimation, i.e. *b* should be somewhere in between *n* and $p + 1$ such that we have sufficient observations for each $\hat{\beta}_i$ and part of them will be significant. This problem is referred as a bias-variance trade-off in choosing the value of *b* (Brunsdon *et al.*, 1996). Since the LWLS estimate for point *i*, $\hat{\beta}_i$, is only an approximation of the true value in the same point, one can consider observations very closed to *i* with the justification that it is reasonable to assume a lower magnitude of difference in the beta values. However, because of the sample size reduction, the standard error of $\hat{\beta}_i$ will increase (i.e. $\hat{\beta}_i$ are not significant). The opposite will occur if we increase the sample size. To solve the problem of choosing an optimal bandwidth value, $b^{opt}$, a cross-validation (CV) method has been widely used, which minimizes the overall residual sum of squares obtained when observation $i$ is deleted (*i.e.* the *target point*). A different and more general approach based on a trade-off between goodness of fit and degrees of freedom is the AIC[6] (Fotheringham *et al.*, 2002, p. 61)

$$b^{opt} = \min_b AIC_c = \min_b 2n \ln(\hat{\sigma}_\varepsilon) + n \ln(2\pi) + n \left\{ \frac{n+tr(S)}{n-2-tr(S)} \right\} \qquad [2.4]$$

where $\hat{\sigma}_\varepsilon$ is the estimated standard deviation of the error term in [2.1] and $tr(S)$ is the trace of the so-called hat matrix S which is in turn a function of the bandwidth. The effective number of degrees of freedom is equal to $n - 2tr(S) - tr(S'S)$, where $S = X(X'W_i X)^{-1} X'W_i$ is the hat matrix (or projection matrix) of $\hat{y}$ on $y$. Then, the effective number of parameters is $p \leq 2tr(S) - tr(S'S) \leq n$ where $p$ is the number of parameters in the corresponding global model. The bandwidth value tends to infinity if $2tr(S) - tr(S'S) = p$ whereas it tends to 0 if $2tr(S) - tr(S'S) = n$.

Another important issue is that the bandwidth value can be defined as fixed or variable/adaptive. Since that adaptive bandwidths suit highly irregular sample configurations and they ensure sufficient (and constant) local information for each local calibration, the adoption of an adaptive bandwidth seems to be preferred to a fixed bandwidth for analyzing spatial data. The choice of an adaptive bandwidth is then useful for two reasons: (i) it guarantees the same number of observations (i.e. the same amount of information) for each local estimation, (ii) it can potentially reduce the edge effects in spatial data analysis. Therefore, in this paper we calculate an adaptive bandwidth following the nearest neighbor approach and we define, for each used kernel weighting function, the optimal variable bandwidth or window size, $b^{opt}$, by using the minimizing the AIC criterion in [2.4]. The bandwidth value (*b*) selects the *b*-smallest distances for each initial local estimation. Each of the elements of the geographical distances are then normalized through the maximum distance, so that it does not matter how far is the most distant point within the optimal neighborhood defined by *b* for each local estimation. In this way, through an appropriate kernel function, the weights that contribute to each local estimation do not depend on the effective position of the points in space. In the following two paragraphs we intend to explain our two-step procedure. The first step (i.e. Section 3) is an algorithm that is able to endogenously identify spatial regimes, whereas the second step (i.e. Section 4) is based on the specification and estimation of two spatial autoregressive models with the identified spatial regimes.

---

[6] Note that the AIC in equation [2.4] is not the same used in the empirical analysis to compare different model specifications. In Section 2 the criterion [2.4] is used to select the initial bandwidth optimal value using the package *GWmodel* in R (Lu et al. 2014), whereas the AIC in the empirical application is the common used for the evaluation of the model fitting.

**3. The first step: an algorithm to deal with unobserved spatial heterogeneity**

In this section we intend to explain our iterative procedure, which enters as a first step of the entire estimation approach. The idea comes from the adaptive weights smoothing (AWS) procedure first proposed by Polzehl and Spokoiny (2000) in the image literature. In particular, we iteratively extend the LWR approach by computing new weights in the main diagonal of $W_i$ at each iteration and then comparing the estimated beta coefficients in [2.3] by using repeated Wald test statistics and smoothing variations of the weights. All the observations that reveal similar beta coefficients (*i.e.* not significant differences in test values) will belong to the same cluster.

The first step starts with the definition of the starting weight vector as a kernel function of both the *distance* between two units in space and the bandwidth value, *i.e.* $w_{ij}^0 = K(d_{ij}; b)$ where $K(.)$ is the kernel function, and 0 represents the "zero" iteration. In order to calculate the optimal value of *b* we consider the AIC in [2.4] for different kernel functions chosen and respect to the model used. We estimate a basic model using a standard kernel weight function, with the optimal window size (i.e. adaptive/variable bandwidth) chosen by minimizing the AIC, and then different coefficients are estimated for each point in the data set. The weight given to observation *j* when estimating the coefficients for observation $i$ is given by $w_{ij}^0$. Initial estimates are then obtained by minimizing equation [2.2] with the initial set of weights $w_{ij}^0$. The approach here proposed is considerably different from that by Andreano et al. (2016) since they do not impose an initial bandwidth value. Moreover, another important difference is based on the calculation of the initial weights. In fact, by imposing a bandwidth value the initial weights are necessary truncated at the maximum distance between the last nearest neighbor point (if an adaptive bandwidth is used). Finally, we added flexibility to the algorithm by considering different kernel functions.

Once the starting weights are defined, we calculate the initial estimates $\hat{\beta}_i^0$ and $\hat{\sigma}_{\varepsilon_i}^{2\,0}$ by using [2.3]. From the second step and until the condition $max|w_{ij}^{l-1} - w_{ij}^l| < \omega\ \forall ij, i \neq j$ holds, with $\omega$ a fixed small value (i.e. 0.0001), we compute *updated weights* for each iteration, *i.e.* $w_{ij}^l\ \forall l$. To this purpose, at the same iteration, say $l$, we compare the local estimates obtained from different spatial units, *i.e.* $\hat{\beta}_i^l$ with $\hat{\beta}_j^l\ \forall ij, i \neq j$ and $\forall l$, by using the following Wald test statistics

$$\chi_{ij}^l = (\hat{\beta}_i^l - \hat{\beta}_j^l)'(\Sigma^l)^{-1}(\hat{\beta}_i^l - \hat{\beta}_j^l) \quad\quad [3.1]$$

where $\Sigma^l$ is the pooled variance-covariance matrix obtained as a weighted average of the two variance-covariance matrices at iteration $l$, *i.e.* $\Sigma^l = [\text{tr}(W_i^l)\Sigma_i^l + \text{tr}(W_j^l)\Sigma_j^l]/[\text{tr}(W_i^l) + \text{tr}(W_j^l)]$, and $(\hat{\beta}_i^l, \hat{\beta}_j^l)$ are $(p+1)$-dimensional column vectors. Note that $\Sigma_i^l = [(X'W_i^l X)^{-1} X'W_i^l][(X'W_i^l X)^{-1} X'W_i^l]'\hat{\sigma}_{\varepsilon_i}^2$ is the variance-covariance matrix of $\hat{\beta}_i^l$ and $\Sigma_j^l$ is the same matrix for $\hat{\beta}_j^l$. This variance-covariance matrix is different from that in Andreano et al. (2016) in order to preserve degrees of freedom.

Each vector $\hat{\beta}_i^l$ for $i = 1, \ldots, n$, at iteration $l$, is obtained from the LWLS problem in [2.3] by only using a subsample of observations whose size is defined by a mix of one of the possible kernel functions and a bandwidth value. If the number of observations is reasonably large in all defined subsamples (which contribute to different beta vector estimations), the approximate multivariate normality is a standard result $\hat{\beta}_i^l \sim N(\beta^l, \Sigma_i^l)$ (see Ibragimov and Müller, 2010 and Bester *et al.*, 2011). Moreover, since the $y_i's$ are assumed to be independently distributed with same variance $\sigma_\varepsilon^2$ (see equation [2.1]), then the estimate of $\Sigma_i^l$ for point $i$ is consistent with a consistent estimate $\hat{\sigma}_{\varepsilon_i}^2$. We then assume that $\hat{\beta}_i^l$ is approximately independent of $\hat{\beta}_j^l$ due to the consistency of $\Sigma_i^l$ and $\Sigma_j^l$ for $i, j = 1, ..., n$. Each statistic $\chi_{ij}^l$ in [3.1] asymptotically follows a $\chi^2$ distribution with $p + 1$ degrees of freedom (which corresponds to the dimension of the column vector $\hat{\beta}_i^l$). The $\chi_{ij}^l \, \forall l$ distance criteria are used as penalties: the larger are their values (i.e. the higher are the distances between pairs of estimated beta vectors) the lower are the weights assigned at each next iteration, and therefore the higher is the probability that units $i$ and $j$ belongs to different groups. The initial assigned weights are then updated in a sense that they decrease if the values of the test statistics are large, and increase otherwise. Therefore, for each observation $i$, we use these test values to calculate a new set of weights, placing higher weight on observations $j$ that have lower test statistics for the difference in coefficients. In order to ensure that the final assigned weights converge to a vector of 0,1 values[7], we perform the convergence procedure quite slowly. As a result we can define the updated weights $w_{ij}^l$ at a generic iteration $l$ by post-multiplying $w_{ij}^0 = K(d_{ij}; b)$ with a kernel that is a function of the above test statistics

$$w_{ij}^l = K(d_{ij}^l; b) K(\chi_{ij}^l; \tau) \qquad [3.2]$$

where $d_{ij}^l = d_{ij}/l$ and it is used to guarantee that further iterations do not decrease in the estimation accuracy, whereas $\tau$ is a parameter that scales the $\chi_{ij}^l$ test statistic. In $d_{ij}^l$ we can also say that the weight of the distance $d_{ij}$ decreases as the number of iteration $l$ increases, so that it increases the weight associated to the value of $\chi_{ij}^l$. In particular, the second Kernel function in equation [3.2] is equal to

$$K(\chi_{ij}^l; \tau) = exp\left(-0.5(\chi_{ij}^l \tau)^2\right).$$

---

[7] At the end of the convergence procedure, the values of the weights which correspond to 1 are assigned to the points which form a cluster and 0 for all the other points. This is equivalent to saying that the procedure allows the weights to converge to a uniform kernel within each region.

Relatively large values of $\tau$ can lead to unstable performance of the procedure, whereas small values reduce the sensitivity to structural changes. The values of $\tau$ can be defined by considering that each $\chi_{ij}^l$ asymptotically follows a $\chi_{p+1}^2$ distribution and then $\tau = 1/t_\alpha(\chi_{p+1}^2)$ where $t_\alpha(\chi_{p+1}^2)$ is the $\alpha - quantile$ of the chi-square distribution (for details see Polzehl and Spokoiny, 2006). Moreover, default values for $\alpha$ for different Kernel functions are provided. In our case we set this parameter equal to 0.001, which is a value that correspond to a value of $\alpha < 0.03$. Finally, we update the resulting weights in [3.2] once again by averaging them with those obtained in the previous step, i.e. $w_{ij}^{l-1}$,

$$\breve{w}_{ij}^l = (1 - \eta)\breve{w}_{ij}^{l-1} + \eta w_{ij}^l \qquad [3.3]$$

where $\eta \in (0,1)$ is a control parameter (also called *memory parameter* in Polzehl and Spokoiny, 2006) which is used to stabilize the convergence procedure. We set $\eta = 0.5$, so that we consider the exact arithmetic mean between the weights at the $(l - 1)$-th and the $l$-th iterations. In this way, the new weights are smoothly or gradually introduced to avoid potential convergence problems due to discontinuities in the iterative process. During the convergence procedure, it is important to provide enough observations for each formed group. Because of the identification problems, groups characterized by a limited number of observations (i.e., less than the number of parameters to be estimated) will be automatically excluded from the analysis and treat as outliers. A final useful definition for the description of the convergence procedure is the *weights variation function*

$$d(w) = \max_{i,j} |\breve{w}_{ij}^{l-1} - w_{ij}^l| \qquad [3.4]$$

which tends to zero as the convergence procedure stabilizes (see Figure 3). This procedure does not impose any restrictions and it is fully adaptive in the sense that no prior information about the spatial structure is required. To summarize, the convergence procedure is based on the following steps

- Define initial weights with a kernel that is a function of distances between pairs of units in space and an optimal bandwidth value (the econometric model must be assumed to be the true one) and use these weights in equation [2.3].

- Once the initial estimates are obtained through equation [2.3], the convergence procedure is ensured by replacing new weights in the main diagonal of equation [2.2] through an updating process of the weights (a modified adapting weights smoothing)

- This updating process is a smoothing process that gradually modifies the initial weights by post-multiplying them with the weights defined through the Wald test statistics (equations [3.1] and [3.2]) and then re-update them through a weighted average with the final weights of the previous iteration, with equation [3.3].

- By setting $\eta = 0.5$ in equation [3.3], we guarantee a gradually introduction of the new weights to be used again in equation [2.3] till convergence, i.e. $d(w) \to 0$.

**4. The second step: spatial autoregressive models with spatial regimes**

It is already well-known that (first-order) autoregressive spatial model with (first-order) autoregressive disturbances (SARARs) models and spatial Durbin models (SDMs) provide general starting points for spatial analysis, since they subsume, in different ways, both spatial autoregressive error (SEM) and spatial autoregressive (SAR) models (LeSage and Pace, 2009). The two spatial models are defined as

$$y = \rho M_1 y + X\beta + u \quad u = \lambda M_2 u + \varepsilon \qquad \varepsilon \sim N(0, \sigma_\varepsilon^2 I_n) \qquad \{SARAR\} \quad [3.5]$$

$$y = \rho M_1 y + X\beta + M_2 \breve{X} \theta + \varepsilon \qquad \varepsilon \sim N(0, \sigma_\varepsilon^2 I_n) \qquad \{SDM\} \quad [3.6]$$

where $X$ is an $n$ by $p+1$ matrix of regressors representing own-region characteristics with a $p+1$ by $1$ vector of parameters $\beta$, $\breve{X}$ is the same matrix without the intercept and $\theta$ is the corresponding $p$ vector of parameters, $\rho$ is the autoregressive parameter, $u$ is an $n$ by $1$ vector of autoregressive error terms with spatial coefficient $\lambda$ and spatial weight matrix $M_2$, $\varepsilon$ is an $n$ by $1$ vector of *i.i.d.* error terms. $M_1$ is the commonly used $n$ by $n$ spatial weighting matrix which can be different or equal to $M_2$. In this paper we set $M_1 = M_2 = M$ and we define $M$ as a *k*-nearest neighbor weight matrix with $k = 10$ and geographical distances. The squared weighting matrix $M$ has the diagonal elements *w<sub>ii</sub>* equal to zero, the rows and columns uniformly bounded as $n$ goes to infinity, and it is then normalized, so that the admissible parameter spaces of $\rho$ and $\lambda$ are known and less than unity in absolute value.

It is worth noting that, although both the weighting matrix used in this section, i.e. $M$, and the one specified in equations [2.2], [2.3] for the local estimations, i.e. $W_i \, \forall i$, are based on a *k*-nearest neighbor criterion, they actually exhibit a different spatial structure. Firstly, each $W_i$ in equations [2.2] and [2.3] is a diagonal *n*-dimensional squared matrix, whereas $M$ not. Secondly, a *k*-nearest neighbor criterion (i.e. based on the optimal adaptive bandwidth value) is only used to define the initial set of weights in $W_i$, so that this is an issue only for the first set of local estimates within the iterative procedure. Finally, there is no need of normalization to ensure admissible parameter spaces of autoregressive coefficients in the case of $W_i$[8].

Recently it has been emerged that SDMs are useful model specifications to overcome the omitted variable problem especially with hedonic house price data (Gerkman, 2012). In order to show how estimating models with spatially varying parameters can produce substantial differences in terms of the autocorrelation effects, in this paper we estimated different global models (GMs): OLS, SAR, SEM, SARAR, and SDM. Those models are then compared with the same models adding spatially varying coefficients, after using the ILWR approach to identify the number of clusters in space. We called these model specifications Endogenous Spatial Regime models (ESRMs). Referring to the global models in [3.5-3.6] we will then have the following ESRMs

$$\begin{bmatrix} y_1 \\ \vdots \\ y_c \end{bmatrix} = \tilde{\rho} M_1 \begin{bmatrix} y_1 \\ \vdots \\ y_c \end{bmatrix} + \begin{bmatrix} X_1 & \cdots & 0 \\ \vdots & \ddots & \vdots \\ 0 & \cdots & X_c \end{bmatrix} \begin{bmatrix} \beta_1 \\ \vdots \\ \beta_c \end{bmatrix} + \begin{bmatrix} u_1 \\ \vdots \\ u_c \end{bmatrix}, \begin{bmatrix} u_1 \\ \vdots \\ u_c \end{bmatrix} = \tilde{\lambda} M_2 \begin{bmatrix} u_1 \\ \vdots \\ u_c \end{bmatrix} + \begin{bmatrix} \varepsilon_1 \\ \vdots \\ \varepsilon_c \end{bmatrix} \quad \{ESR\text{-}SARAR\} \quad [3.7]$$

$$\begin{bmatrix} y_1 \\ \vdots \\ y_c \end{bmatrix} = \tilde{\rho} M_1 \begin{bmatrix} y_1 \\ \vdots \\ y_c \end{bmatrix} + \begin{bmatrix} X_1 & \cdots & 0 \\ \vdots & \ddots & \vdots \\ 0 & \cdots & X_c \end{bmatrix} \begin{bmatrix} \beta_1 \\ \vdots \\ \beta_c \end{bmatrix} + M_2 \begin{bmatrix} \overline{X}_1 & \cdots & 0 \\ \vdots & \ddots & \vdots \\ 0 & \cdots & \overline{X}_c \end{bmatrix} \begin{bmatrix} \tilde{\theta}_1 \\ \vdots \\ \tilde{\theta}_c \end{bmatrix} + \begin{bmatrix} \varepsilon_1 \\ \vdots \\ \varepsilon_c \end{bmatrix} \quad \{ESR\text{-}SDM\} \quad [3.8]$$

where, in both cases, $\varepsilon_j \sim N\left(0, \sigma_{\varepsilon_j}^2 I_{n_j}\right)$, $j = 1, \dots, c$, and $c$ is the unknown number of clusters in space. After having defined these unknown clusters through the first step procedure, we proceed with the estimation of models in [3.7] and [3.8], which can be interpreted as spatial autoregressive models with structural breaks in space (also known in the literature as spatial regimes).

---

[8] The weights in the diagonal of $W_i$ have the role, at first, of selecting the observations that contribute to local estimations trough local weighted least squares estimations, whereas the weights in $M$ permits, trough the reduced form of the model, the expansion of the infinite number of cross-sectional effects in the entire system (i.e. *global spillover effects*).

**5. Data sets and results**

In order to detect the presence of unobserved spatial heterogeneity in the form of spatial regimes and to prove the usefulness of the above two-step approach, we use *baltimore* and *house* data sets[9] which summarizes the information on house sales prices in Baltimore, MD, and Lucas County, Ohio, respectively (see Table 1). Baltimore data set was first analyzed by Dubin (1992) who emphasized the problem of spatially varying parameters. Because of the computational burden of our algorithm in the first step we selected the central subsample of the cross-sectional house data set in 1993 (see Figure 1).

INSERT TABLE 1 HERE

INSERT FIGURE 1 HERE

---

[9] For details see the *spdep* package in R (Bivand, 2014).

Tables 2 and 3 provide GMs and ESRMs estimates for Baltimore data set. The bandwidth value, based on the AIC criterion (expr. [2.4]), corresponds to 33 nearest neighbors, $b_{knn}^{AIC} = 33$, whereas the total number of observations are 211. The first step identified two homogeneous clusters ($c = 2$) with subsamples $n_1 = 101$ and $n_2 = 110$, mostly dividing the central and the northwest areas from the south and the northeast ones (Figure 2(a)). Considering AIC values, ESRMs are always to be preferred over GMs and in both situations the SAR specification seems to fit the data best. Then, in this case positive spatial neighbor effects seem to persist despite the presence of two spatial regimes. However, if we see the results of SDM with spatial regimes (SDM-SRs) we can confirm the hypothesis that most of the spatial autocorrelation is induced by the omitted variable problem of not considering the lagged covariates. In this case the presence of spatial regimes has a "pure" informative role: the SDM-SR specification (with lowest AIC) is more useful than the others since it can be used to guide zone-specific policies.

INSERT FIGURE 2 HERE

INSERT TABLE 2 HERE

INSERT TABLE 3 HERE

Tables 4 and 5 show GMs and ESRMs estimates for house data set, instead. The bandwidth value corresponds to 19 nearest neighbors, $b_{knn}^{AIC} = 19$, over 382 number of observations. The first step identified 4 homogeneous clusters ($c = 4$) with subsamples $n_1 = 57$, $n_2 = 141$, $n_3 = 58$, $n_4 = 126$ (Figure 2(b)). Conclusions are a bit different from those obtained for Baltimore: not only ESRMs provide better results in terms of AIC values, but also spatial spillover effects seem to be attenuated by the presence of the spatial regimes. This is particularly clear in the comparison between a global SDM and an SDM-SRs where the autocorrelation coefficient goes down from 0.655 to 0.248. In this case the additional information of the identified spatial regimes makes us able to split the apparent contagion (in Anselin's words) into the true contagion (the spillover effects), the part due to the lagged covariates, and the part due to the presence of spatial heterogeneity.

INSERT TABLE 4 HERE

INSERT TABLE 5 HERE

It is then interesting to note that estimating spatial models with spatial regimes (endogenously identified with the first step of our procedure) can be not only useful to account for both spatial dependence and spatial heterogeneity giving more flexibility to our models, but also to provide zone-specific economic results. Considering for instance Baltimore results (Tables 2-3), we can observe that some regressors in ESRMs are highly statistically significant in the first zone (i.e., central and northwest areas) whereas not in the second (i.e., the south and the northeast areas). Therefore, for the same model a significant global parameter estimate should be mainly due to the statistical significance of the same parameter in central and northwest areas.

It is well known in the literature that house prices do not necessarily change at a uniform rate throughout an urban area. Several papers that attempt to smoothly describe a spatial price variation or to identify housing submarkets within cities have been published (see e.g. McMillen, 2013, for a recent exhaustive review). The identification of regimes in housing markets allows us to identify a set of discrete neighborhoods with constant coefficients within each subarea and discrete changes across boundaries. Therefore, this procedure could be viewed as a solution for the identification of the potential number of submarkets (see e.g. Watkins, 2001) within an urban area (e.g. a city) especially when no prior knowledge on the geographical structure is available.

INSERT TABLE 6 HERE

INSERT FIGURE 3 HERE

Table 6 shows the number of iterations of the algorithm and the time required to ensure the convergence (see Figure 3). Although the computational time seems to be still reasonable as long as the number of observations are lower than 400, we can note that the time is dramatically increased by an augmentation of only around 100 of observations, and the number of iterations are twice as much those required for *baltimore* data set. Conversely, the number of explanatory variables seems not to change the time required by our algorithm. Finally, in Figure 3 we can see how the convergence procedure works: after an initial period of "instability", the weights tend to zero as the number of iterations increased. The most significant and rapid decrease is between the interval of iterations ($5^{th}$-$15^{th}$) for baltimore and ($15^{th}$-$30^{th}$) for house.

## 6. Conclusions and future developments

Spatial heterogeneity is one form of unobserved heterogeneity, which leads to model misspecification problems. Geographically weighted regressions or geoadditive models are surely promising alternatives to account for spatially varying parameters, in which the parameter surface is not assumed to be constant over space. However, economic studies usually suggest a form of spatially varying parameters, which can be interpreted as the presence of discrete spatial heterogeneity (or spatial regimes). In this paper we then define a two-step approach able to identify these regimes in a first step and to simultaneously estimate spatial dependence and heterogeneity in a second step.

By using two house price data sets, we found that our procedure was able to identify clusters and that spatial econometric models estimated on that partition were to be preferred. Most of the results showed that spatial neighbor effects are present in our data sets but in a reduced form, and probably part of the neighborhood effects is not only due to an omitted variable problem but also to the unobserved spatial heterogeneity. This firstly suggested that accounting for both spatial dependence and some form of spatial heterogeneity (e.g. the spatial regimes, with the use of some algorithm-based approach to identify them) is a fundamental step. The additional information derived from the identified spatial regimes makes us able to split the apparent contagion into three parts: the true contagion (the spillover effects), the lagged covariates (which justifies the use of the spatial Durbin model), and the presence of spatial heterogeneity. Moreover, zone-specific economic results can be provided with these more flexible spatial models.

As expected, its main drawback is the computational time required for the convergence procedure, which we suppose can become infeasible with very large data sets. The need to overcome the ILWR computational problems is straightforward. Unobserved spatial heterogeneity is still a problem in spatial econometrics. Many author are still trying to test the presence of spatial heterogeneity, but the prospect of considering algorithm-based methods is becoming a successful way to follow in order to solve practical problems with several data sets and to consider higher flexibility in our models.

The first step of our two-step procedure does not impose any restrictions and is fully adaptive in the sense that no prior information about the spatial structure is required. It can be easily extended to any other model which can be locally estimated by weighted least squares, and with different definition of the weights (e.g. economic definitions rather than mere geographic distances). Moreover, two possible and interesting modifications of this algorithm can be the one that permits the simultaneous consideration of both spatial dependence and local marginal effect estimation inside the iterative procedure and the extension to panel data sets to also deal with time variations. These will be accounted for in future works.

**Acknowledgments**

A preliminary version of the paper has been presented at the 55$^{th}$ European Regional Science Association (ERSA) Congress held in Lisbon, 25-28 August, 2015. We thank all the participants, especially those in our spatial econometrics and regional economic modelling session.

Watkins, C. A. (2001) The definition and identification of housing submarkets, Environment and Planning A 33, 2235-2253.

# Tables and Figures

**Table 1**
Variable description.

|  | Variable | Description |
|---|---|---|
| Baltimore data set | price | price of the house. |
|  | dwell | dummy variable. 1 for the single-family houses. |
|  | nbath | number of baths. |
|  | patio | dummy variable. 1 if there is a patio. |
|  | firepl | dummy variable. 1 if there is a fireplace. |
|  | ac | dummy variable. 1 if there is an air conditioner. |
|  | bment | number of basements. |
|  | gar | number of garages. |
|  | citcou | dummy variable. 1 for houses located in Baltimore County. |
|  | lotsz | lot size measured in hundred square feet. |
| House data set | price | price of the house. |
|  | yrbuilt | year of the building. |
|  | TLA | transitional living area. |
|  | baths | number of baths. |
|  | halfbaths | number of half baths. |
|  | garagesqft | garage size measured in square feet. |
|  | lotsize | lot size measured in square feet. |

**Table 2**
Parameter estimates of the global models in the baltimore data set.

| baltimore data set | GLOBAL MODELS | | | | | | | | | |
|---|---|---|---|---|---|---|---|---|---|---|
| Coefficients | OLS | | SAR | | SEM | | SARAR | | SDM | |
| intercept | 2.463 |  | -6.173 | . | 3.674 |  | -6.125 | . | 22.920 | * |
| dwell | 8.171 | *** | 7.336 | *** | 9.223 | *** | 7.547 | *** | 8.422 | *** |
| nbath | 7.653 | *** | 6.839 | *** | 7.854 | *** | 6.960 | *** | 6.226 | *** |
| patio | 9.948 | *** | 8.443 | *** | 7.661 | ** | 8.343 | *** | 6.985 | ** |
| firepl | 11.910 | *** | 9.740 | *** | 8.868 | *** | 9.592 | *** | 9.024 | *** |
| ac | 8.350 | *** | 6.970 | ** | 7.974 | *** | 7.061 | ** | 8.704 | *** |
| bment | 3.508 | *** | 3.380 | *** | 3.469 | *** | 3.405 | *** | 3.011 | *** |
| gar | 5.638 | ** | 5.456 | *** | 4.866 | ** | 5.408 | *** | 5.047 | *** |
| citcou | 12.537 | *** | 9.426 | *** | 13.130 | *** | 9.609 | *** | 11.603 | *** |
| lotsz | 0.043 | * | 0.034 | * | 0.037 | * | 0.034 | * | 0.039 | ** |
| M_dwell |  |  |  |  |  |  |  |  | -13.645 | . |
| M_nbath |  |  |  |  |  |  |  |  | -10.522 | * |
| M_patio |  |  |  |  |  |  |  |  | 12.765 | . |
| M_firepl |  |  |  |  |  |  |  |  | 21.653 | ** |
| M_ac |  |  |  |  |  |  |  |  | 5.415 |  |
| M_bment |  |  |  |  |  |  |  |  | -4.580 |  |
| M_gar |  |  |  |  |  |  |  |  | 6.016 |  |
| M_citcou |  |  |  |  |  |  |  |  | -3.775 |  |
| M_lotsz |  |  |  |  |  |  |  |  | 0.118 | . |
| rho | - |  | 0.334 | *** | - |  | 0.324 | *** | 0.062 |  |
| lambda | - |  | - |  | 0.490 | ** | 0.050 |  | - |  |
| AIC | 1681.14 |  | 1664.14 |  | 1672.57 |  | 1666.092 |  | 1651.608 |  |

Signif. codes: 0 '***' 0.001 '**' 0.01 '*' 0.05 '.' 0.1 ' ' 1

**Table 3**
Parameter estimates of models with spatial regimes in the baltimore data set based on the partition in Figure 2(a).

| baltimore data set | MODELS WITH SPATIAL REGIMES | | | | | | | | |
|---|---|---|---|---|---|---|---|---|---|
| Coefficients | OLS | | SAR | | SEM | | SARAR | | SDM | |
| **intercept1** | -4.997 | | -16.409 | *** | -5.274 | | -16.670 | *** | -194.163 | * |
| **intercept2** | 12.042 | . | 0.206 | | 11.374 | * | -0.602 | | -16.976 | |
| **dwell1** | 6.314 | * | 6.004 | * | 8.635 | ** | 5.269 | . | 5.551 | * |
| **dwell2** | 8.222 | ** | 7.570 | ** | 7.864 | ** | 7.440 | ** | 6.329 | * |
| **nbath1** | 5.463 | ** | 5.406 | ** | 5.546 | ** | 5.330 | ** | 4.292 | ** |
| **nbath2** | 9.818 | ** | 8.527 | *** | 8.413 | ** | 8.635 | ** | 8.005 | ** |
| **patio1** | 11.699 | *** | 10.328 | *** | 8.924 | ** | 10.696 | *** | 12.514 | *** |
| **patio2** | 0.560 | | 1.010 | | 2.744 | | 0.392 | | 1.960 | |
| **firepl1** | 15.794 | *** | 12.420 | *** | 12.203 | *** | 12.733 | *** | 15.219 | *** |
| **firepl2** | 5.643 | | 6.087 | . | 4.934 | | 6.441 | * | 5.329 | . |
| **ac1** | 8.157 | * | 7.620 | ** | 7.999 | ** | 7.517 | ** | 3.514 | |
| **ac2** | 6.611 | * | 4.800 | . | 5.794 | * | 4.478 | | 3.110 | |
| **bment1** | 7.303 | *** | 7.885 | *** | 7.853 | *** | 7.699 | *** | 8.082 | *** |
| **bment2** | 1.959 | | 1.403 | | 1.503 | | 1.390 | | 0.987 | |
| **gar1** | 7.453 | *** | 7.342 | *** | 6.782 | *** | 7.402 | *** | 2.128 | |
| **gar2** | 0.433 | | -0.085 | | -0.479 | | 0.090 | | 0.324 | |
| **citcou1** | 18.453 | *** | 14.134 | *** | 20.252 | *** | 13.149 | *** | 14.575 | *** |
| **citcou2** | 8.948 | ** | 6.211 | * | 8.810 | ** | 5.954 | * | 12.333 | *** |
| **lotsz1** | 0.042 | . | 0.033 | . | 0.033 | . | 0.034 | . | 0.032 | |
| **lotsz2** | 0.046 | . | 0.036 | | 0.041 | . | 0.035 | | 0.034 | |
| **M_dwell1** | | | | | | | | | -106.986 | ** |
| **M_dwell2** | | | | | | | | | 33.649 | |
| **M_nbath1** | | | | | | | | | 166.857 | ** |
| **M_nbath2** | | | | | | | | | 19.496 | |
| **M_patio1** | | | | | | | | | 286.654 | *** |
| **M_patio2** | | | | | | | | | -27.744 | |
| **M_firepl1** | | | | | | | | | -218.294 | *** |
| **M_firepl2** | | | | | | | | | 25.074 | |
| **M_ac1** | | | | | | | | | 94.830 | *** |
| **M_ac2** | | | | | | | | | -29.819 | |
| **M_bment1** | | | | | | | | | -187.329 | *** |
| **M_bment2** | | | | | | | | | 58.434 | |
| **M_gar1** | | | | | | | | | 108.945 | * |
| **M_gar2** | | | | | | | | | 32.013 | |
| **M_citcou1** | | | | | | | | | 0.235 | |
| **M_citcou2** | | | | | | | | | -0.417 | |
| **M_lotsz1** | | | | | | | | | 6.357 | . |
| **M_lotsz2** | | | | | | | | | 2.192 | |
| **rho** | - | | 0.327 | *** | - | | 0.354 | *** | -0.086 | |
| **lambda** | - | | - | | 0.494 | ** | -0.139 | | - | |
| **AIC** | 1662.762 | | 1646.486 | | 1655.166 | | 1648.300 | | 1621.114 | |

Signif. codes: 0 '***' 0.001 '**' 0.01 '*' 0.05 '.' 0.1 ' ' 1

**Table 4**

Parameter estimates of the global models in the house data set.

| house data set | GLOBAL MODELS | | | | | | | |
|---|---|---|---|---|---|---|---|---|
| **Coefficients** | OLS | | SAR | | SEM | | SDM | |
| **intercept** | -9.99E+05 | *** | -5.53E+05 | *** | -4.39E+05 | *** | -3.05E+05 | * |
| **yrbuilt** | 5.12E+02 | *** | 2.72E+02 | *** | 2.22E+02 | *** | 2.10E+02 | *** |
| **TLA** | 1.98E+01 | *** | 1.64E+01 | *** | 2.07E+01 | *** | 1.96E+01 | *** |
| **baths** | 1.03E+04 | ** | 1.42E+04 | *** | 1.46E+04 | *** | 1.52E+04 | *** |
| **halfbaths** | 1.33E+04 | *** | 4.98E+03 | *** | 5.01E+03 | *** | 5.19E+03 | *** |
| **garagesqft** | 2.13E+01 | *** | 1.40E+01 | *** | 1.19E+01 | *** | 1.20E+01 | ** |
| **lotsize** | 2.00E+00 | *** | 9.56E-01 | *** | 1.03E+00 | *** | 9.97E-01 | *** |
| **M_yrbuilt** | | | | | | | -5.52E+01 | |
| **M_TLA** | | | | | | | -1.85E+01 | *** |
| **M_baths** | | | | | | | -1.06E+04 | |
| **M_halfbaths** | | | | | | | 5.84E+03 | |
| **M_garagesqft** | | | | | | | 3.02E+00 | |
| **M_lotsize** | | | | | | | 9.02E-01 | |
| **rho** | - | | 0.580 | *** | - | | 0.655 | *** |
| **lambda** | - | | - | | 0.831 | *** | - | |
| **AIC** | 8498.578 | | 8302.719 | | 8289.333 | | 8269.549 | |

Signif. codes:  0 '***' 0.001 '**' 0.01 '*' 0.05 '.' 0.1 ' ' 1

**Table 5**

Parameter estimates of models with spatial regimes in the house data set based on the partition in Figure 2(b).

| house data set | MODELS WITH SPATIAL REGIMES | | | | | | | |
|---|---|---|---|---|---|---|---|---|
| **Coefficients** | OLS | | SAR | | SEM | | SDM | |
| **intercept1** | -5.68E+05 | * | -5.08E+05 | ** | -7.53E+05 | *** | 1.92E+08 | . |
| **intercept2** | -3.29E+05 | | -6.27E+05 | *** | -4.93E+05 | *** | -6.49E+07 | . |
| **intercept3** | 4.58E+05 | | -1.52E+05 | | -2.09E+05 | . | 3.85E+07 | |
| **intercept4** | -5.80E+05 | * | -8.60E+05 | *** | -6.40E+05 | *** | 8.93E+07 | |
| **yrbuilt1** | 2.97E+02 | ** | 2.54E+02 | * | 3.91E+02 | *** | 3.89E+02 | *** |
| **yrbuilt2** | 4.72E+02 | *** | 3.24E+02 | *** | 2.62E+02 | *** | 3.05E+02 | *** |
| **yrbuilt3** | 5.78E+01 | | 7.67E+01 | | 1.12E+02 | | 1.18E+02 | |
| **yrbuilt4** | 5.84E+02 | *** | 4.29E+02 | *** | 3.21E+02 | *** | 2.62E+02 | *** |
| **TLA1** | 1.36E+01 | . | 1.23E+01 | * | 1.30E+01 | * | 1.29E+01 | * |
| **TLA2** | 1.15E+01 | ** | 1.17E+01 | *** | 1.32E+01 | *** | 1.77E+01 | *** |
| **TLA3** | 9.56E+00 | | 8.51E+00 | . | 7.92E+00 | . | 7.27E+00 | |
| **TLA4** | 2.92E+01 | *** | 2.39E+01 | *** | 2.82E+01 | *** | 2.68E+01 | *** |
| **baths1** | 2.67E+04 | ** | 2.28E+04 | ** | 2.01E+04 | ** | 1.57E+04 | * |
| **baths2** | 3.93E+03 | | 2.99E+03 | | 1.15E+03 | | -6.26E+02 | |
| **baths3** | 6.70E+02 | | 8.31E+02 | | 2.83E+03 | | 3.42E+03 | |
| **baths4** | 2.07E+03 | | 1.13E+04 | * | 1.11E+04 | *** | 1.38E+04 | *** |
| **halfbaths1** | 3.06E+03 | | 1.94E+03 | | -1.53E+03 | | 3.68E+03 | |
| **halfbaths2** | 6.45E+03 | | 2.68E+03 | | 3.87E+02 | | -2.08E+02 | |
| **halfbaths3** | -3.01E+03 | | -1.70E+03 | | -4.83E+03 | | -4.72E+03 | |
| **halfbaths4** | 1.64E+04 | *** | 8.19E+03 | *** | 6.19E+03 | *** | 9.39E+03 | *** |
| **garagesqft1** | 1.08E+01 | | 1.59E+01 | | 2.09E+01 | . | 3.92E+00 | |
| **garagesqft2** | 1.39E+01 | * | 1.17E+01 | * | 1.00E+01 | * | 9.55E+00 | * |
| **garagesqft3** | 1.63E+01 | . | 1.28E+01 | . | 9.84E+00 | | 9.85E+00 | |
| **garagesqft4** | 1.80E+01 | ** | 1.04E+01 | . | 1.28E+01 | * | 8.76E+00 | . |
| **lotsize1** | 1.32E+00 | ** | 1.03E+00 | * | 1.08E+00 | ** | 1.29E+00 | ** |
| **lotsize2** | 6.62E-01 | | 5.49E-01 | | 9.33E-01 | * | 9.36E-01 | * |
| **lotsize3** | -1.83E-02 | | 9.73E-02 | | 1.24E+00 | | 8.70E-01 | |
| **lotsize4** | 2.14E+00 | *** | 1.02E+00 | * | 1.35E+00 | *** | 1.21E+00 | *** |
| **M_yrbuilt1** | | | | | | | -3.92E+02 | ** |
| **M_yrbuilt2** | | | | | | | 7.77E+01 | ** |
| **M_yrbuilt3** | | | | | | | -9.27E+00 | |
| **M_yrbuilt4** | | | | | | | -9.42E+01 | ** |
| **M_TLA1** | | | | | | | 2.08E+04 | |
| **M_TLA2** | | | | | | | -2.42E+05 | *** |
| **M_TLA3** | | | | | | | 8.55E+03 | |
| **M_TLA4** | | | | | | | 1.01E+05 | * |
| **M_baths1** | | | | | | | 3.76E+05 | *** |
| **M_baths2** | | | | | | | 1.29E+05 | ** |
| **M_baths3** | | | | | | | 1.79E+04 | |
| **M_baths4** | | | | | | | 1.10E+05 | *** |
| **M_halfbaths1** | | | | | | | -7.62E+02 | ** |
| **M_halfbaths2** | | | | | | | 1.42E+02 | * |
| **M_halfbaths3** | | | | | | | 1.17E+02 | |
| **M_halfbaths4** | | | | | | | -5.84E+01 | |
| **M_garagesqft1** | | | | | | | 3.36E+01 | ** |
| **M_garagesqft2** | | | | | | | -4.90E-01 | |
| **M_garagesqft3** | | | | | | | -7.41E+00 | |
| **M_garagesqft4** | | | | | | | 7.22E-01 | |
| **M_lotsize1** | | | | | | | -2.06E+02 | . |
| **M_lotsize2** | | | | | | | 6.92E+01 | *** |
| **M_lotsize3** | | | | | | | -4.16E+01 | |
| **M_lotsize4** | | | | | | | -9.66E+01 | |
| **rho** | - | | 0.444 | *** | - | | 0.248 | *** |
| **lambda** | - | | - | | 0.791 | *** | - | |
| **AIC** | 8356.899 | | 8252.417 | | 8246.132 | | 8174.823 | |

Signif. codes: 0 '***' 0.001 '**' 0.01 '*' 0.05 '.' 0.1 ' ' 1

**Table 6**
Computational details.

| Data sets | total # observations | # parameters for each iteration | # iterations | CPU time (in sec.) |
|---|---|---|---|---|
| **Baltimore** | 211 | 9 | 96 | 1683.12 |
| **House** | 328 | 6 | 182 | 13069.14 |

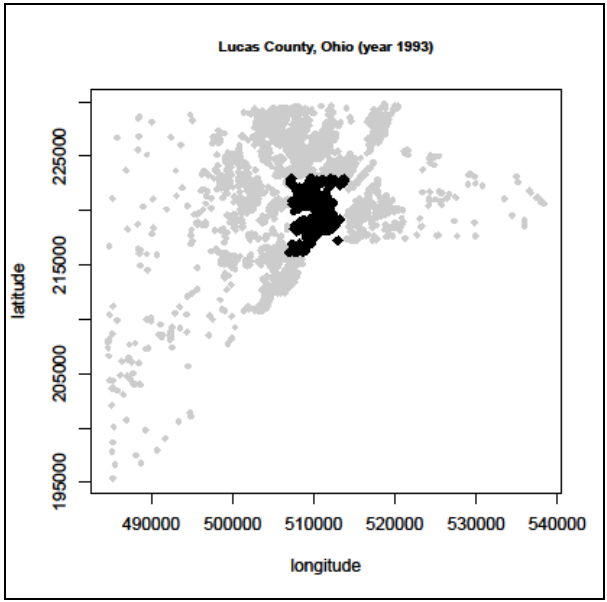

**Figure 1.** Selected data from House data set in 1993 year.

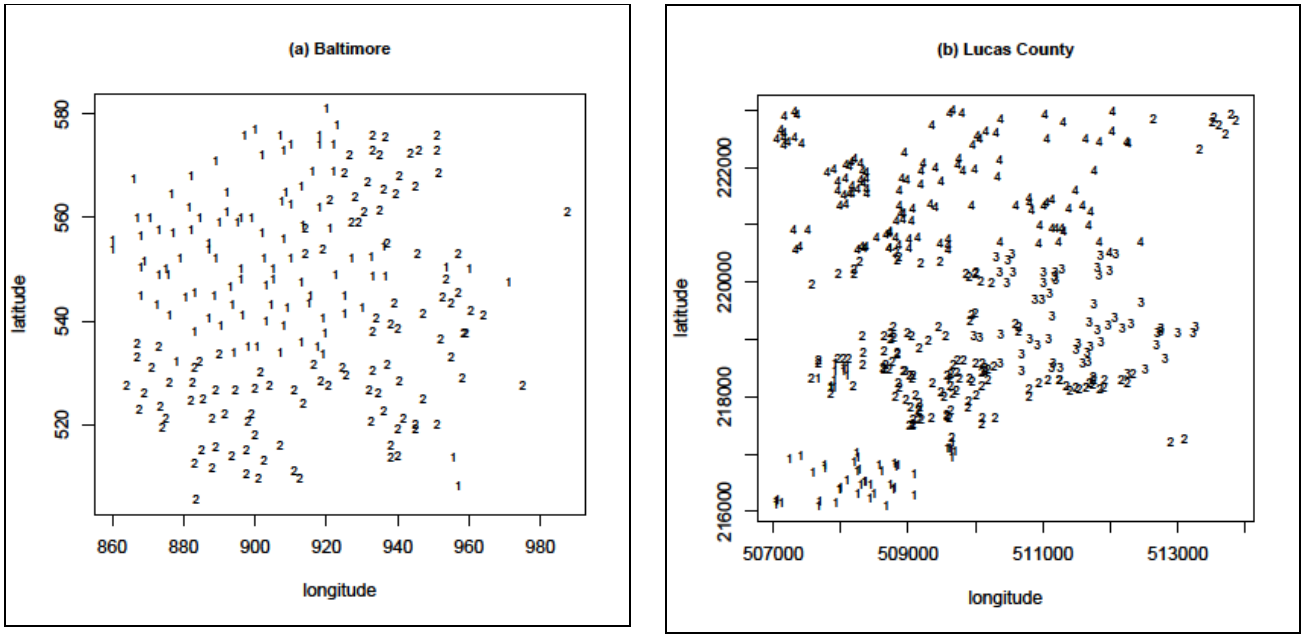

**Figure 2.** Spatial regimes in (a) *baltimore* and (b) *house* data sets. The number of regimes are $c = 2$ and $c = 4$ respectively, whereas the whole sample sizes are $n = 211$ and $n = 382$ respectively.

Note: Gaussian kernel functions for initial and updated weights are used. Adaptive bandwidths are based on the AIC in equation [2.4]: (a) $b_{knn}^{AIC} = 33$, (b) $b_{knn}^{AIC} = 19$. $\tau = 0.001$ is fixed.

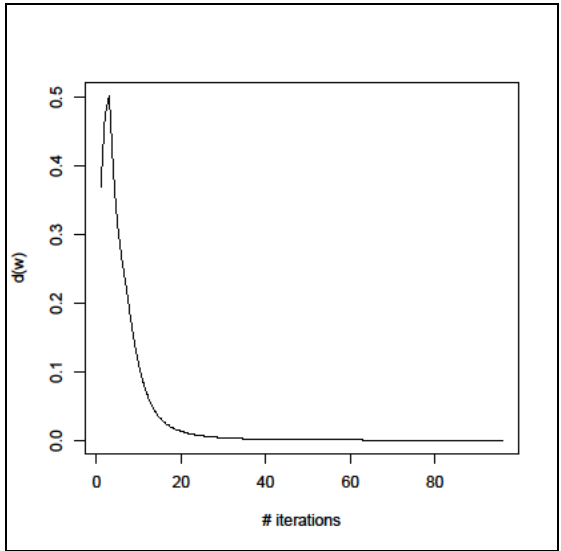 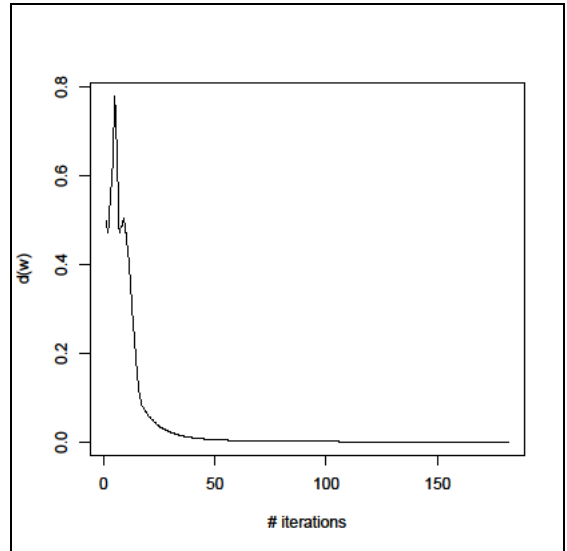

(a)     (b)

**Figure 3.** Weights variation functions from (a) *baltimore* and (b) *house* data sets.